\documentclass[pdflatex,sn-mathphys-num]{sn-jnl}

\usepackage{booktabs}
\usepackage{siunitx}
\usepackage{graphicx}%
\usepackage{multirow}%
\usepackage{amsmath,amssymb,amsfonts}%
\usepackage{amsthm}%
\usepackage{mathrsfs}%
\usepackage[title]{appendix}%
\usepackage{xcolor}%
\usepackage{textcomp}%
\usepackage{manyfoot}%
\usepackage{booktabs}%
\usepackage{algorithm}%
\usepackage{algorithmicx}%
\usepackage{algpseudocode}%
\usepackage{listings}%

\theoremstyle{thmstyleone}%
\theoremstyle{thmstyletwo}%

\theoremstyle{thmstylethree}%

\begin{document}

\title[WaveletEM]{Frequency-Aware Dual-Stream Learning for Balanced Realism and Fidelity in Electron Microscopy Imaging}


\author[1]{\fnm{Longmi} \sur{Gao}}\email{gaolongmi@nuaa.edu.cn}
\author[1]{\fnm{Zhengkai} \sur{Zhao}}\email{zhengkai.zhao@nuaa.edu.cn}
\author*[1]{\fnm{Pan} \sur{Gao}}\email{Pan.Gao@nuaa.edu.cn}
\author[2]{\fnm{Manoranjan} \sur{Paul}}\email{mpaul@csu.edu.au}

\affil*[1]{\orgdiv{College of Artificial Intelligence}, \orgname{Nanjing University of Aeronautics and Astronautics}, \orgaddress{\city{Nanjing}, \postcode{211106}, \country{China}}}
\affil[2]{\orgdiv{School of computing, Mathematics, and Engineering}, \orgname{Charles Sturt University}, \orgaddress{\state{NSW}, \country{Australia}}}


\abstract{Electron microscopy enables nanoscale cellular visualization but faces a trade-off between imaging resolution and acquisition speed. Existing learning-based methods rely on single-stream architectures that struggle to balance perceptual realism and quantitative fidelity, either over-smoothing details or generating unrealistic hallucinations. This work introduces a frequency-adaptive dual-stream architecture to resolve this conflict. Using discrete wavelet transform, we decompose images into low-frequency structures and high-frequency details, then employ a conditional diffusion model for realistic global synthesis and a transformer network for precise detail recovery. Experiments on the EMDiffuse dataset show the method achieves superior LPIPS and resolution ratio, substantially outperforming existing approaches. The method also shows strong generalization across diverse biological samples, supporting fast and reliable electron microscopy imaging for structural biology and nanotechnology applications. The source code and associated dataset are publicly available to facilitate further research.
}




\maketitle

\section{Introduction}\label{sec1}

Electron microscopy (EM), with its nanoscale resolution capabilities, serves as an indispensable tool across structural biology, materials science, nanotechnology, and semiconductor research~\cite{TransmissionEM,EM1,EM2,EM3,EM4,EM5}. As a critical branch of EM~\cite{EM6,EM7}, scanning electron microscopy (SEM) has continually evolved through advances in high-resolution imaging~\cite{EM8,EM9}, aberration correction~\cite{EM10,EM11}, and artificial intelligence integration~\cite{EM12}, offering unprecedented insights into nanoscale structures. However, the physical imaging process is inherently constrained by the ``eternal triangle of compromise"~\cite{EM13}, where optimizing image resolution, electron beam intensity, or acquisition speed inevitably degrades the others~\cite{Fang2021}. Within this triad, spatial resolution\footnote{Resolution is defined as the minimum distance between two point emitters (or image features), which depends on the imaging system's diffraction limit, detector sampling, and signal-to-noise characteristics.} is paramount, as it determines the limit of discernible structural details. Specifically, acquiring high-resolution SEM images necessitates prolonged scanning durations and elevated electron beam intensities, which not only exacerbate the accumulation of random noise but also significantly increase the risk of irradiation damage. Consequently, to bypass these physical hardware limitations, developing advanced computational methods to enhance image resolution and fidelity under restricted sampling frameworks has emerged as a critical research imperative. This endeavor lies at the intersection of microscopy and visual computing, as it demands sophisticated image restoration and reconstruction techniques to computationally recover high-quality visual information from degraded acquisitions. More broadly, deep learning-based biomedical image analysis is increasingly demonstrating translational impact across diverse clinical applications, from stroke risk prediction to disease screening~\cite{Jiang2025Deep,Meng2025Noninvasive}, reinforcing the importance of developing faithful and robust reconstruction methods for scientific imaging.

Deep learning has emerged as a transformative force in EM image enhancement~\cite{tip3}, yet the field currently faces a fundamental dilemma. We distinguish three interrelated concepts throughout this paper: \textit{perceptual realism} (the visual naturalness of generated textures), \textit{quantitative fidelity} (pixel-wise accuracy measured by PSNR and SSIM), and \textit{biological faithfulness} (the minimum resolvable distance between adjacent features in the frequency domain). While deterministic pixel-wise methods (e.g., PSSR (Point-Scanning Super-Resolution)~\cite{Fang2021}) and generative models (e.g., EMDiffuse~\cite{Lu2024}) have achieved significant strides, they struggle to balance perceptual realism with quantitative fidelity. We attribute this limitation to the inherent spectral bias of single-stream neural networks, which fail to simultaneously optimize contradictory frequency characteristics. Specifically, deterministic reconstruction models\cite{attention1,CNN1} (e.g., CNNs or Transformers trained with pixel-wise losses) inherently minimize pixel-wise error, prioritizing low-frequency modes at the cost of high-frequency loss, resulting in over-smoothed artifacts that compromise biological realism~\cite{oversmooth}, as shown in Figure~\ref{fig:1}. Conversely, pure diffusion models focus on matching data distributions, often generating plausible high-frequency textures that lack global consistency, leading to stochastic hallucinations that degrade structural fidelity.

To overcome this dichotomy, it is essential to operate in a domain that effectively disentangles these conflicting components. The Discrete Wavelet Transform~\cite{WT} provides an ideal mathematical basis for this purpose by decomposing the image into multi-scale subbands. This property is uniquely suited for EM imagery, as it allows us to spatially localize and independently process global morphological structures (low-frequency) and fine-grained textural details (high-frequency). Recent work has also demonstrated the value of frequency-domain enhancement in biomedical image analysis~\cite{Wen2025Lightweight}. Consequently, this decomposition enables the application of specialized optimization strategies tailored to the distinct characteristics of each frequency component.

In this work, we introduce WaveletEM, a frequency-aware dual-stream architecture designed to resolve the realism-fidelity trade-off. To our knowledge, WaveletEM is among the first approaches to explicitly assign diffusion-based structural synthesis and transformer-based detail reconstruction to separate wavelet frequency bands for electron microscopy enhancement. Our core strategy, Frequency-Aware Structural Decomposition, leverages the DWT to explicitly separate an EM image into `semantic structure' (low-frequency) and `textural details' (high-frequency). This decomposition enables a collaborative reconstruction approach: we employ a Low Frequency Conditional Diffusion Model (LFCDM) to generatively synthesize biologically plausible structural skeletons, while simultaneously utilizing a High Frequency Transformer-Based Model (HFTBM) to deterministically recover fine-grained textures with high precision. By synergizing these specialized branches, our method effectively integrates generative realism with deterministic fidelity. Extensive experiments on multiple electron microscopy datasets demonstrate that our method achieves superior performance in both quantitative metrics and qualitative visual assessment, notably outperforming existing approaches in terms of perceptual quality, resolution improvement, and structural integrity.
Our contributions can be summarized as follows:
\begin{figure}[]
  \includegraphics[width=\textwidth]{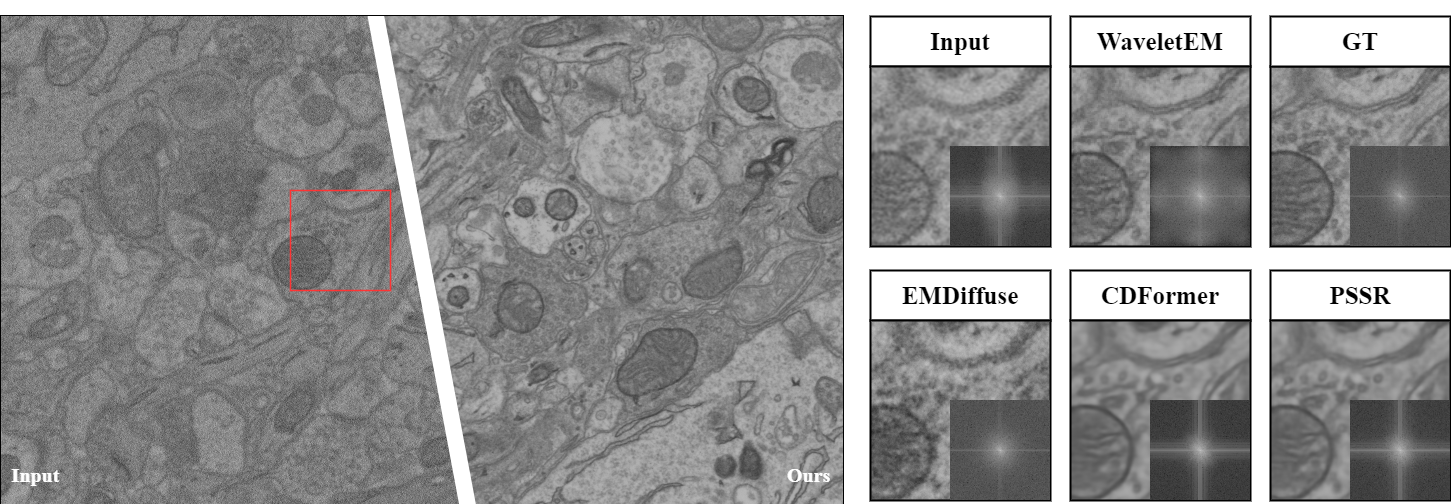}
  \caption{Visual comparison of our proposed WaveletEM against two established electron microscopy imaging algorithms (PSSR and EMDiffuse) and a recent image enhancement method (CDFormer) on the Mouse Brain Cortex dataset. Each subfigure includes its corresponding Fourier power spectrum in the lower-right corner, and each pixel represents 3.3 nm. While previous methods (e.g., CDFormer and PSSR) struggle to reconstruct multi-frequency domain information, or fail to generate high-clarity and realistic texture details (e.g., EMDiffuse), WaveletEM effectively balances multi-frequency fidelity and structural authenticity, producing clearer, more detailed reconstructions.}
  \label{fig:1}
\end{figure}
\begin{itemize}
\item We propose WaveletEM, a dual-stream architecture that leverages the discrete wavelet transform to explicitly disentangle frequency components, effectively resolving the inherent conflict between perceptual realism and quantitative fidelity in EM imaging.
\item We design a novel Frequency-Aware Structural Decomposition strategy that synergizes a Low Frequency Conditional Diffusion Model for authentic structural synthesis and a High Frequency Transformer-Based Model for precise detail restoration.
\item Extensive experiments demonstrate that WaveletEM achieves superior performance, particularly in LPIPS and resolution ratio, while maintaining competitive PSNR and SSIM, and generalizes robustly across diverse biological samples.
\end{itemize}

\section{Preliminary}\label{sec2}

\subsection{2D Discrete Wavelet Transform}
In this work, we employ the 2D discrete wavelet transform with Haar wavelets~\cite{Haar}. While other bases like Daubechies  or Biorthogonal were considered, we selected Haar for its superior balance of performance and efficiency in our context. Many alternative wavelets, lacking properties like symmetry or orthogonality, can introduce artifacts or are computationally more demanding, proving less suitable for precise image recovery tasks. The orthogonality of the Haar wavelet provides a stable and non-redundant mathematical basis, which is fundamental for the effective independent processing in our dual-stream architecture. Furthermore, our choice aligns with the current trend in numerous image reconstruction studies~\cite{WTConv,sadat_litevae_2024,phung_wavelet_2023,DiffLL} that leverage the simple yet powerful Haar wavelet~\cite{2DWT1,2DWT2,2DWT3}. In the following, we present the mathematical foundation of 2D-DWT from a convolution perspective~\cite{WTConv}.

Given an input image $X$, the 2D discrete Haar wavelet transform can be interpreted as convolving the image with four fixed filters with a stride of 2, generating four subbands with halved spatial resolution. Each subband captures distinct frequency information: low-frequency approximation, and high-frequency details in horizontal, vertical, and diagonal directions.
Formally, the decomposition process can be represented using four convolution filters:
\begin{equation}
f_{LL} = \frac{1}{2}
\begin{bmatrix}
\phantom{-}1 & \phantom{-}1 \\
\phantom{-}1 & \phantom{-}1
\end{bmatrix}, \quad
f_{LH} = \frac{1}{2}
\begin{bmatrix}
\phantom{-}1 & -1 \\
\phantom{-}1 & -1
\end{bmatrix}, \quad
f_{HL} = \frac{1}{2}
\begin{bmatrix}
\phantom{-}1 & \phantom{-}1 \\
-1 & -1
\end{bmatrix}, \quad
f_{HH} = \frac{1}{2}
\begin{bmatrix}
\phantom{-}1 & -1 \\
-1 & \phantom{-}1
\end{bmatrix}
\label{eq:1}
\end{equation}
The forward wavelet transform can then be expressed as a convolution operation:
\begin{equation}
    [X_{LL}, X_{LH}, X_{HL}, X_{HH}] = \text{Conv}([f_{LL}, f_{LH}, f_{HL}, f_{HH}], X)
    \label{eq:2}
\end{equation}
where $X_{LL}$ represents the low-frequency approximation coefficient, while $X_{LH}$, $X_{HL}$, $X_{HH}$ capture horizontal, vertical, and diagonal high-frequency details, respectively.

The inverse wavelet transform, which reconstructs the original image from the wavelet coefficients, can similarly be formulated as a deconvolution operation:
\begin{equation}
X = \text{Deconv} \left( [f_{LL}, f_{LH}, f_{HL}, f_{HH}], [X_{LL}, X_{LH}, X_{HL}, X_{HH}] \right).
\end{equation}

This decomposition provides a natural multi-scale representation that effectively separates structural content from high-frequency details, making it particularly suitable for our electron microscopy image enhancement task.

\subsection{Related Work}
Deep learning has revolutionized image enhancement tasks, particularly super-resolution (SR) and denoising, evolving from CNNs to Transformers and diffusion models.
CNN-based approaches established the foundation for learning-based enhancement. For SR, SRCNN~\cite{dong2015SRCnn} pioneered this direction, while VDSR~\cite{kim2016VDSR} introduced deeper networks with residual learning. In denoising, DnCNN~\cite{zhang2017DnCNN} employed residual learning to outperform traditional methods. Perceptual quality was significantly advanced by SRGAN~\cite{ledig2017SRGAN}, which generated more realistic high-frequency details through adversarial training.
Subsequently, Transformer-based architectures emerged to address the limited receptive fields of convolutional models. SwinIR~\cite{liang2021swinir} leveraged hierarchical Transformers with shifted windows for SR, while Restormer~\cite{zamir2022restormer} enhanced efficiency with specialized self-attention for high-resolution image denoising.
Diffusion models mark the latest advancement in image enhancement. SRDiff~\cite{li2022srdiff} iteratively refines SR images through conditional diffusion, while DDRM~\cite{kawar2022DDRM} demonstrates superior denoising via score-based diffusion. Recently, DiffLL~\cite{DiffLL} integrated wavelet-based frequency-domain operations into diffusion models for low-light enhancement, opening new avenues for frequency-aware diffusion in image restoration.

In the context of electron microscopy, deep learning has substantially advanced both image acquisition and quantitative analysis across a wide spectrum of tasks: robust denoising and reconstruction from low-dose or fast-scan acquisitions~\cite{EM3}, super-resolution restoration of subcellular structures beyond hardware limits~\cite{Fang2021}, automated detection and characterization of nanoscale particles~\cite{EM12}, learning-based microstructural segmentation for materials quantification~\cite{EM2}, and macromolecular structure determination in cryo-EM~\cite{EM1}. Among these, image denoising and super-resolution are particularly critical for addressing the fundamental trade-off between acquisition speed and image quality. PSSR~\cite{Fang2021} introduced the ``crappifier approach", which deliberately degrades high-quality images by simultaneously introducing noise and reducing spatial resolution, enabling simultaneous supersampling and denoising. However, such deterministic reconstruction models tend to produce overly smooth images, reducing overall realism. To address this, EMDiffuse~\cite{Lu2024} emerged as a pioneering diffusion-based framework for EM image enhancement, leveraging the probabilistic nature of diffusion models to generate high-resolution, noise-reduced images while preserving structural details. EMDiffuse implements an ensemble strategy to stabilize generation, yet its inherent stochasticity can occasionally compromise image fidelity, particularly in low-frequency signal fitting.

Our work is further motivated by concurrent progress in image restoration beyond electron microscopy. SAT-Net~\cite{SATNet} shows that structure-aware attention mechanisms benefit medical image enhancement; SCNet~\cite{SCNet} demonstrates the effectiveness of dual-branch architectures for strong noise removal; CCM-Net~\cite{Wen2025CCMNet} highlights the importance of artifact suppression and quality-aware analysis in medical imaging; and UTDM~\cite{UTDM} illustrates the generalization of transformer-based diffusion models to complex degradations. These studies collectively validate the potential of combining dual-stream design with diffusion and transformer modules, which we adapt to the frequency-decomposed setting of EM imaging.

In light of these limitations, we propose WaveletEM, a wavelet-based dual-stream architecture. By leveraging the multi-frequency analysis capabilities of wavelet transforms, our approach simultaneously enhances image realism and fidelity.

\section{Method}\label{sec3}
In this section, we elaborate on our proposed WaveletEM architecture. We begin by presenting the overall dual-stream design. Following this, we introduce the two core components of our architecture: the Low Frequency Conditional Diffusion Model and High Frequency Transformer-Based Model. Finally, we discuss our training strategy with separate supervision.
\begin{figure*}[t]
    \centering
    \includegraphics[width=\textwidth]{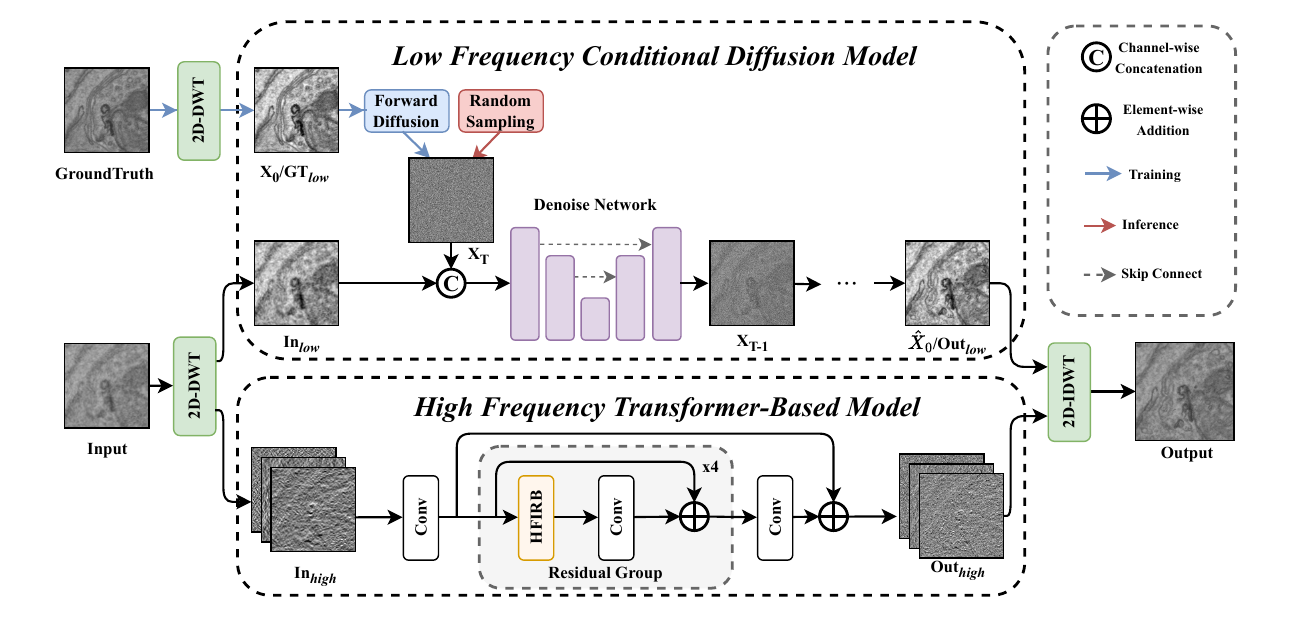}
    \caption{Overall architecture of our proposed WaveletEM. Color coding denotes the operational scope of each component: blue arrows and blocks indicate training-only paths, red elements represent inference-only paths, and black elements are shared across both training and inference. During inference, only the forward pass through the black and red pathways is required; the blue-coded training components (e.g., the noise addition forward diffusion process) are not executed.}
    \label{model:1}
\end{figure*}
\subsection{Network Architecture}
As illustrated in Figure~\ref{model:1}, our methodology is built upon three fundamental components: a wavelet transform for frequency decomposition and reconstruction, a Low Frequency Conditional Diffusion Model for authentic structural synthesis, and a High Frequency Transformer-Based Model for fine-detail restoration. Specifically, given an input electron microscopy image, we first employ the 2D Discrete Wavelet Transform (2D-DWT) to decompose the signal into distinct frequency domains. This operation yields four subbands: the low-frequency approximation $X_{LL}$ and three high-frequency detail subbands ($X_{LH}$, $X_{HL}$, $X_{HH}$). We denote the low-frequency component as $In_{low} = X_{LL}$, and the high-frequency components jointly as $In_{high} = \{X_{LH}, X_{HL}, X_{HH}\}$. These components are subsequently processed via two specialized parallel streams. In the generative branch, $In_{low}$ serves as a structural condition; it is concatenated with a pure Gaussian noise latent $X_{T}$ and fed into the LFCDM to synthesize the denoised structural output $Out_{low}$. Simultaneously, the three high-frequency subbands $In_{high}$ ($X_{LH}$, $X_{HL}$, $X_{HH}$) are concatenated channel-wise and jointly processed by the HFTBM, which performs deterministic reconstruction to recover fine-grained details, resulting in $Out_{high}$. Ultimately, the enhanced components $Out_{low}$ and $Out_{high}$ are fused via the 2D Inverse Discrete Wavelet Transform (2D-IDWT) to yield the final high-resolution image.

\subsection{Low Frequency Conditional Diffusion Model}
In the ground truth data distribution, numerous high-frequency detail information exists, as illustrated by the Fourier power spectrum in Figure~\ref{fig:1}. For such non-smooth prediction problems, conventional CNN or transformer-based models inevitably lose these high-frequency details, evident in the Fourier power spectra of PSSR and CDFormer~\cite{CDFormer} shown in Figure~\ref{fig:1}.  While GAN-based approaches offer an alternative, they often suffer from unstable training processes and limited capability in handling geometric transformations~\cite{GAN1}. In contrast, diffusion models exhibit greater stability in training and improved performance in image generation and restoration tasks due to their distinctive diffusion-based training and inference schemes~\cite{GAN2}.

Our conditional diffusion model operates by gradually adding Gaussian noise to the low-frequency components of the EM images and then learning to reverse this process. For a low-frequency image $X_0$, we define a forward diffusion process that progressively adds noise according to:
\begin{equation}
q\left(X_{t} \mid X_{t-1}\right)=\mathcal{N}\left(X_{t} ; \sqrt{1-\beta_{t}} X_{t-1}, \beta_{t} \mathbf{I}\right)
\end{equation}

The forward process can be expressed in a closed form for any arbitrary timestep
$t$, allowing direct sampling from \( q(x_t \mid x_0) \)
\begin{equation}
X_t = \sqrt{\overline{\alpha}_t}\, X_0 + \sqrt{1 - \overline{\alpha}_t}\,\epsilon
\end{equation}
where $\alpha_t = 1 - \beta_t$, $ \overline{\alpha}_t = \prod_{i=1}^t \alpha_i$, 
and $\epsilon \sim \mathcal{N}(0, I)$.
This one-step noise addition efficiently enables training at random timesteps.

Subsequently, the noisy image $X_t$ is input into DDPM~\cite{DDPM} for single-step denoising (reverse process), as shown below:
\begin{equation}
    X_{t-1} 
= \frac{1}{\sqrt{\alpha_t}}
  \Bigl(
    X_t - \frac{1 - \alpha_t}{\sqrt{1 - \overline{\alpha}_t}} \,\epsilon_\theta(X_t, t, c)
  \Bigr)
+ \sigma_tz
\end{equation}
where \(\epsilon_\theta\) is our noise prediction network conditioned on 
the low-frequency component \(c\), \(z \sim \mathcal{N}(0, I)\) is an additional 
noise term, and \(\sigma_t\) controls the stochasticity of the reverse process. 
This formulation enables our model to progressively denoise the image while 
incorporating conditional information from the low-frequency component.

We optimize our model using a simplified objective:
\begin{equation}
\label{eq:loss1}
Loss_{diff}
= \mathbb{E}_{X_0,\,c,\,\epsilon,\,t}
\bigl[
||\epsilon - \epsilon_\theta(X_t, t, c)||^2
\bigr].
\end{equation}

By leveraging this conditional diffusion model for the low-frequency branch, our method effectively captures the underlying structure of EM images while preserving crucial details that CNN, transformer, and GAN-based approaches often fail to maintain.

\subsection{High Frequency Transformer-Based Model}
\begin{figure}[t]
    \centering
    \includegraphics[width=0.7\textwidth]{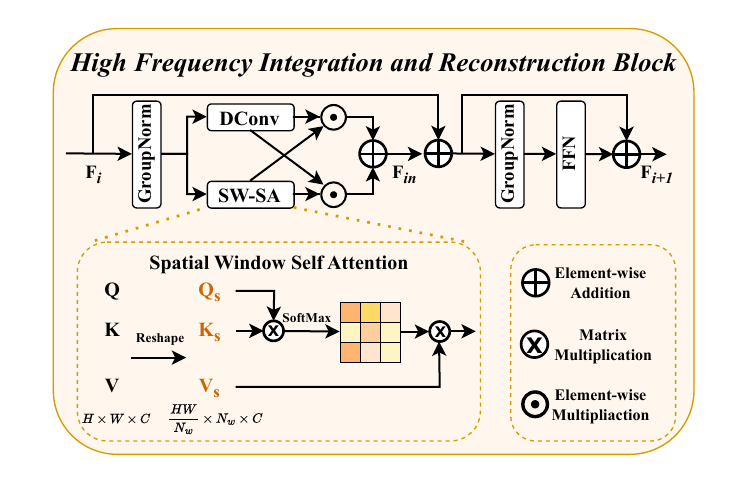}  
    \caption{Details of High Frequency Integration and Reconstruction Block (HFIRB).}
    \label{model:2}
\end{figure}

To enhance the high-frequency information and fidelity in reconstructed EM images, we propose a High Frequency Transformer-Based Model that specifically targets the restoration of three high-frequency coefficients, as illustrated in Figure~\ref{model:1}. The HFTBM architecture consists of a series of convolutional layers interspersed with our High Frequency Integration and Reconstruction Block (HFIRB), forming a residual group structure that effectively captures and refines multi-scale features.

The HFIRB module, depicted in Figure~\ref{model:2}, leverages a dual-path architecture to integrate complementary strengths of CNNs and Transformers. To enhance feature representation capabilities, we employ depth-wise separable convolutions (DConv)~\cite{DWconv} in one path and Spatial Window Self-Attention (SW-SA)~\cite{SWSA} in the parallel path. This design enables efficient processing of both local and global contextual information while maintaining computational efficiency.

The core of our HFIRB module is the Interflow mechanism~\cite{SWSA}, which facilitates bidirectional feature exchange between the CNN and Transformer branches. Given the input feature $F_i$ , the parallel processing and feature exchange can be formulated as:

\begin{equation}F_{in} = \widetilde{\mathbf{\mu}} (SWSA(F_i)) * DConv(F_i) + \widetilde{\mathbf{\sigma}} (DConv(F_i)) * SWSA(F_i)\end{equation}

Where $\widetilde{\mathbf{\mu} }$ and $\widetilde{\mathbf{\sigma}}$ denote the spatial distiller and channel distiller, respectively. The processed features are then refined through a Feed-Forward Network (FFN):
\begin{equation}
F_{i+1} = FFN(GroupNorm(F_{in}))
\end{equation}

This cross-stream information flow allows for mutual enhancement of features, where the local precision of CNNs complements the global reasoning of Transformers. The combination of Group Normalization and Feed-Forward Network further refines the feature representations by adaptively adjusting the feature statistics before generating the final output.

For training our HFTBM model, we employ the L1 loss function to optimize the reconstruction of high-frequency components. The L1 loss function is defined as:
\begin{equation}
    Loss_{tf} = \mathbb{E}\left[\left|Out_{high} - GT_{high}\right|\right]
    \label{eq:loss2}
\end{equation}

By combining these components, the HFTBM effectively reconstructs high-frequency coefficients with enhanced detail preservation, which is crucial for revealing fine structures in electron microscopy images.

\subsection{Training Strategy with Separate Supervision}
To ensure training stability and optimal convergence, we employ a separate supervision strategy that effectively decouples the optimization of the dual streams. By avoiding a unified end-to-end loss, we mitigate potential gradient interference between the heterogeneous diffusion-based and deterministic branches. Specifically, we tailor the objective functions to the distinct characteristics of each frequency domain: an $L_2$ loss is applied to the low-frequency branch ($Loss_{diff}$) to enforce global structural consistency, while an $L_1$ loss is utilized for the high-frequency branch ($Loss_{tf}$) to robustly recover fine details and preserve edge sharpness against noise.

\section{Experiment}\label{sec4}

\subsection{Experimental Setting}

{\bfseries Efficiency and Implementation Protocol.}
WaveletEM is implemented using PyTorch 2.0 with CUDA 11.8. All experiments were conducted on a single NVIDIA RTX 3090 GPU (24 GB) with mixed-precision (FP16) training enabled. The model was trained for $1 \times 10^5$ iterations with batch size 8, using the Adam optimizer~\cite{kingma2017adammethodstochasticoptimization} with an initial learning rate of $1\times 10^{-4}$ and a step decay schedule (halved every $2.5\times10^4$ iterations). Input images were processed at 256$\times$256 pixels for denoising and upscaled from 128$\times$128 to 256$\times$256 for super-resolution. The diffusion model was trained with 2,000 diffusion steps; inference used 1,000 DDPM sampling steps. All inference times reported in this paper are per-image (measured as total batch time divided by batch size, with batch size 10 for timing measurements), excluding data loading and preprocessing, and were averaged over 100 forward passes after a 10-iteration warm-up on the same GPU. For a fair comparison, all competing methods were retrained from scratch on the same EMDiffuse training set using their official implementations and recommended settings, and evaluated on the identical held-out test split.

{\bfseries Datasets.}
Our experiments were conducted on the EMDiffuse dataset~\cite{lu_chixiang_2023_10205819}, which comprises electron microscopy images from diverse biological samples. The mouse brain cortex dataset served as our primary dataset for training and evaluation, while mouse liver, bone marrow, heart, and HeLa cell datasets were utilized for transfer learning experiments.
The images were acquired with varying dwell times ranging from 2s to 130s, representing different noise levels to train a robust model and comprehensively evaluate performance across noise levels. Each anatomical region contains images captured at 20,000× magnification (used for super-resolution tasks) and 40,000× magnification (used for denoising tasks) at different imaging speeds.
\begin{table}[t]
\centering
\caption{Quantitative comparisons on the Mouse Brain Cortex Dataset. The best results are highlighted in \textcolor{red}{red}, and the second-best in \textcolor{blue}{blue}.}
\label{table:1}

\begin{tabular}{l c ccccc}
\hline
Method & Reference & PSNR $\uparrow$ & SSIM $\uparrow$ & LPIPS $\downarrow$ & FSIM $\uparrow$ & Res Ratio $\uparrow$ \\
\hline
\multicolumn{7}{c}{\textbf{Denoising}} \\
\hline
PSSR~\cite{Fang2021} & Nat.Meth., 2021 & \textcolor{red}{28.865} & \textcolor{red}{0.6866} & 0.0411 & 0.9484 & 0.7632 \\
SwinIR~\cite{liang2021swinir} & ICCVW, 2021 & 27.047 & 0.5538 & 0.0541 & 0.9493 & 0.6998 \\
SRDiff~\cite{li2022srdiff} & Neurocomputing, 2022 & 25.478 & 0.5541 & 0.0198 & 0.9501 & 0.9056 \\
SR3~\cite{SR3} & TPAMI, 2023 & 26.864 & 0.5461 & \textcolor{blue}{0.0161} & 0.9482 & 0.9666 \\
CDFormer~\cite{CDFormer} & CVPR, 2024 & 27.612 & 0.5949 & 0.0456 & \textcolor{blue}{0.9507} & 0.7126 \\
EMDiffuse~\cite{Lu2024} & Nat.Comm., 2024 & 23.466 & 0.5145 & 0.0164 & 0.9479 & \textcolor{blue}{0.9925} \\
\midrule
Ours & -- & \textcolor{blue}{28.327} & \textcolor{blue}{0.6264} & \textcolor{red}{0.0133} & \textcolor{red}{0.9508} & \textcolor{red}{1.8579} \\
\hline
\end{tabular}

\vspace{0.5em}

\begin{tabular}{l c ccccc}
\hline
Method & Reference & PSNR $\uparrow$ & SSIM $\uparrow$ & LPIPS $\downarrow$ & FSIM $\uparrow$ & Res Ratio $\uparrow$ \\
\hline
\multicolumn{7}{c}{\textbf{Super-Resolution (2×)}} \\
\hline
PSSR~\cite{Fang2021} & Nat.Meth., 2021 & \textcolor{red}{27.853} & \textcolor{red}{0.6277} & 0.0475 & \textcolor{blue}{0.9506} & 0.6626 \\
SwinIR~\cite{liang2021swinir} & ICCVW, 2021 & 25.985 & 0.5366 & 0.0646 & 0.9478 & 0.6104 \\
SRDiff~\cite{li2022srdiff} & Neurocomputing, 2022 & 24.269 & 0.5374 & 0.0197 & 0.9492 & 0.9245 \\
SR3~\cite{SR3} & TPAMI, 2023 & 25.819 & 0.5181 & \textcolor{blue}{0.0183} & 0.9412 & \textcolor{blue}{1.0113} \\
CDFormer~\cite{CDFormer} & CVPR, 2024 & 26.431 & 0.5716 & 0.0548 & 0.9502 & 0.6301 \\
EMDiffuse~\cite{Lu2024} & Nat.Comm., 2024 & 23.312 & 0.4787 & 0.0214 & 0.9487 & 0.9825 \\
\midrule
Ours & -- & \textcolor{blue}{27.237} & \textcolor{blue}{0.5881} & \textcolor{red}{0.0163} & \textcolor{red}{0.9508} & \textcolor{red}{1.9259} \\
\hline
\end{tabular}

\end{table}
{\bfseries Metrics.}
To comprehensively assess reconstruction quality, we employ PSNR and SSIM~\cite{SSIM} to measure pixel-level and structural fidelity, alongside LPIPS~\cite{lpips} and FSIM~\cite{fsim} to evaluate perceptual quality and feature similarity.

Resolution Ratio. This metric quantifies resolution enhancement by comparing the resolutions of the generated output and the ground truth. Resolution is derived via image decorrelation analysis~\cite{resolution}. 
The method calculates the Pearson correlation between the image's Fourier transform $I(k)$ and a masked, normalized version of itself $I_{n}(k)$, using a circular mask $M(k,r)$ of increasing radius $r$ to yield a decorrelation function:
\begin{equation}
\label{eq:dr}
d(\mathbf{r})
=
\frac{
  \displaystyle
  \iint {Re}\bigl\{I(\mathbf{k})\,I_n^*(\mathbf{k})\,M(\mathbf{k},\mathbf{r})\bigr\}
  \,dk_x\,dk_y
}{
  \displaystyle
  \sqrt{
    \iint \bigl|I(\mathbf{k})\bigr|^2\,dk_x\,dk_y
    \;\;
    \iint \bigl|I_n(\mathbf{k})\,M(\mathbf{k},\mathbf{r})\bigr|^2\,dk_x\,dk_y
  }
}
\end{equation}
where $k = [k_{x},k_{y}]$ denotes Fourier space coordinates.

To isolate high-frequency details and determine the limits of resolution, the above analysis is repeated on a series of high-pass filtered images. First, the original image is high-pass filtered several times with different filters. For the original image and each filtered image, the complete $d(r)$ curve is calculated and the location of the local maximum (peak) on the curve is found. We denote the peak found in the $i$-th analysis as $[r_{i},A_{i}]$, where $r_{i}$ is the frequency position of the peak and $A_i$ is the amplitude of the peak. The resolution is then determined from the maximum decorrelation peak position $r_{i}$ found across these filtered images:
\begin{equation}
\label{eq:resolution}
\mathrm{Resolution}
=
\frac{2 \times \text{pixel size}}
     {\max\bigl(r_0, \dots, r_{N_g}\bigr)}
\end{equation}
where $N_{g}$ represents the number of high-pass filters.

Finally, the Resolution Ratio is defined as the ratio of the GT resolution to the Output resolution:
\begin{equation}
\label{eq:resolution_ratio}
Resolution\:Ratio = \frac{GT\:Resolution}{Output\:Resolution} 
\end{equation}

\begin{figure*}[t]
    \centering
    \includegraphics[width=\textwidth]{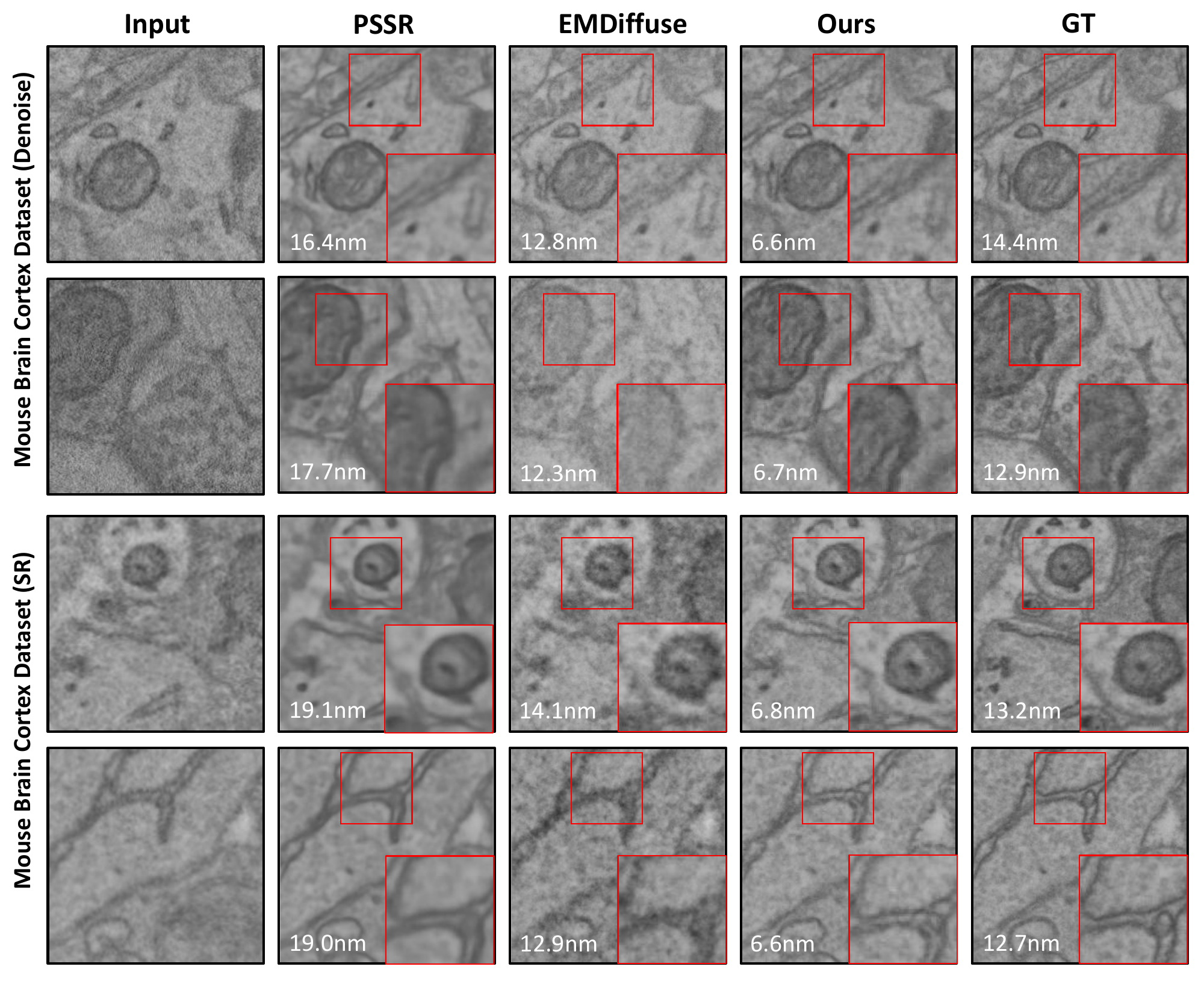}
    \caption{Visualization results on the Mouse Brain Cortex Dataset, demonstrating both denoising and super-resolution performance. The pixel size of all images in this figure is 3.3 nm (a physical acquisition parameter determined by the microscope hardware). The resolution value indicated below each image is the decorrelation resolution (in nm) computed via Fourier image decorrelation analysis (Eq.~\ref{eq:resolution}), which quantifies the finest structural detail recoverable in the frequency domain. These two quantities are distinct: pixel size is a fixed hardware constant, whereas decorrelation resolution is a content-dependent metric that varies across methods and reflects their high-frequency recovery capability.}
    \label{fig:2}
\end{figure*}

\begin{table}[t]
\centering
\caption{Comparison with EMDiffuse in terms of parameter count, computational cost, and inference time (batch size = 10).}
\label{table:4}

\setlength{\tabcolsep}{6pt} 
\begin{tabular}{l S[table-format=2.2] S[table-format=2.2] S[table-format=1.2]}
\toprule
Method & {Params (M)} & {FLOPs (T)} & {Time (s)} \\
\midrule
EMDiffuse & 15.16 & 45.01 & 8.53 \\
WaveletEM & 15.45 & 11.26 & 2.54 \\
\bottomrule
\end{tabular}
\end{table}

\begin{table}[t]
\centering
\caption{Generalization and transferability on the Mouse Brain Cortex dataset. BM denotes Bone Marrow.}
\label{table:gengeate}

\setlength{\tabcolsep}{5pt}
\begin{tabular}{l S[table-format=2.2] S[table-format=1.2] S[table-format=1.3] 
                S[table-format=2.2] S[table-format=1.2] S[table-format=1.3]}
\toprule
& \multicolumn{3}{c}{Without Finetune} & \multicolumn{3}{c}{With Finetune} \\
\cmidrule(lr){2-4} \cmidrule(lr){5-7}
Dataset & {PSNR} & {SSIM} & {LPIPS} & {PSNR} & {SSIM} & {LPIPS} \\
\midrule
BM    & 25.38 & 0.60 & 0.016 & 26.58 & 0.63 & 0.013 \\
Heart & 26.73 & 0.65 & 0.014 & 28.55 & 0.71 & 0.012 \\
Liver & 27.15 & 0.70 & 0.015 & 27.97 & 0.74 & 0.013 \\
HeLa  & 24.23 & 0.49 & 0.071 & 26.39 & 0.53 & 0.048 \\
\bottomrule
\end{tabular}
\end{table}

\subsection{Comparison with Existing Methods}
{\bfseries Comparison with EM imaging algorithm.}
We first benchmark WaveletEM against specialized EM methods: EMDiffuse and PSSR. As presented in Table~\ref{table:1}, WaveletEM achieves the performance with a PSNR of 28.32 dB and SSIM of 0.6264, surpassing EMDiffuse by significant margins (+4.861 dB in denoising). Qualitative results in Figure~\ref{fig:2} further highlight this advantage: while EMDiffuse outputs suffer from residual noise and blurred boundaries, WaveletEM effectively recovers intricate tissue patterns and sharp membrane edges. Regarding PSSR, although it maintains competitive pixel-level metrics due to its pixel-wise optimization objective, it exhibits severe ``over-smoothing" artifacts that compromise biological realism. In contrast, WaveletEM achieves a roughly three-fold improvement in LPIPS (0.0133 vs.\ 0.0411) and doubles the resolution compared to PSSR, confirming its superiority in resolving fine-grained subcellular structures.

Furthermore, as illustrated in Table~\ref{table:4}, WaveletEM demonstrates remarkable computational efficiency compared to EMDiffuse. With comparable parameter counts, our model reduces computational complexity (FLOPs) by approximately 75\%, yielding a roughly 3.4-fold reduction in inference time. WaveletEM processes images in just over two seconds—a significant improvement over EMDiffuse's inference time. Moreover, from the perspective of effective imaging throughput, conventional high-quality SEM acquisition in the EMDiffuse dataset requires dwell times ranging from 100 s to 130 s per frame to achieve the signal-to-noise ratio of the ground-truth images~\cite{lu_chixiang_2023_10205819}. In contrast, the WaveletEM pipeline—acquiring at a reduced dwell time (e.g., 2 s, the fastest setting in the dataset) followed by model inference (2.18 s per image at 1000 sampling steps, Table~\ref{table:2})—reduces the total per-image time from over 100 s to approximately 4.2 s. Comparing the conventional long-exposure acquisition time (\textasciitilde130 s) to the inference time alone (\textasciitilde2.18 s) yields an approximate 60$\times$ acceleration in the computational component of the imaging workflow. It should be noted that direct inference speed comparisons between heterogeneous method categories (e.g., single-pass deterministic models vs.\ iterative diffusion models) are less informative due to fundamental differences in their operational paradigms. For reference, among diffusion-based approaches, WaveletEM achieves approximately 3.4$\times$ faster inference than EMDiffuse (2.54 s vs.\ 8.53 s per image, Table~\ref{table:4}) while delivering superior reconstruction quality.

{\bfseries Comparison with Image Enhancement algorithm.}
We further evaluate our method against leading general-purpose restoration models: the transformer-based CDFormer and the diffusion-based SR3. WaveletEM outperforms SR3 with PSNR and SSIM improvements of 1.463 dB and 0.0803 in denoising tasks. More importantly, our method demonstrates a decisive advantage in perceptual metrics and detail recovery. As illustrated in Table~\ref{table:1}, WaveletEM significantly surpasses CDFormer in LPIPS and achieves a Resolution Ratio exceeding 250\%. These results confirm that by leveraging our Frequency-Aware Structural Decomposition to explicitly disentangle and optimize frequency components, WaveletEM effectively overcomes the fidelity-realism trade-off that limits conventional single-stream enhancement algorithms.

\begin{figure*}[t]
    \centering
    \includegraphics[width=\textwidth]{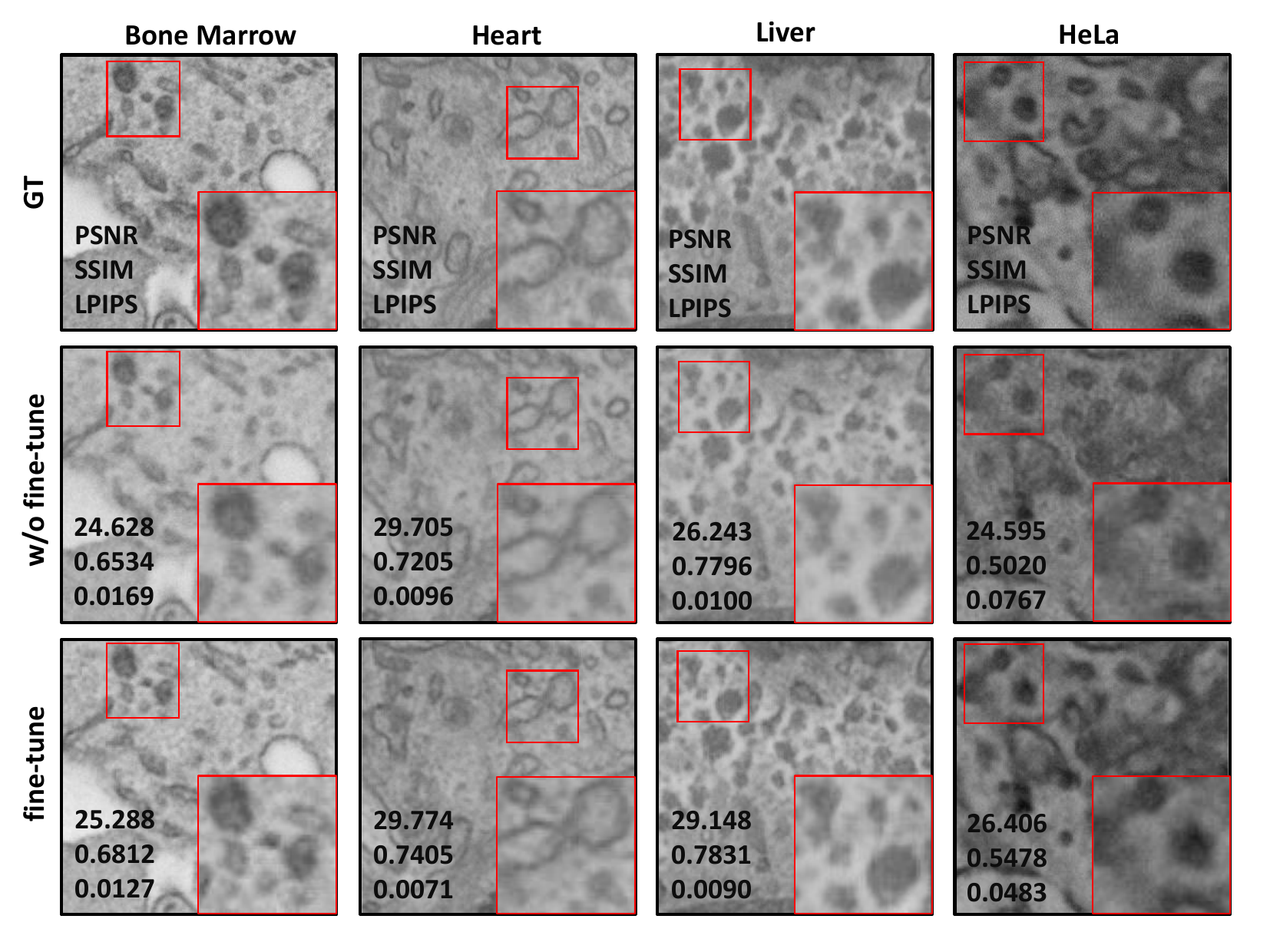}
    \caption{Visualization of WaveletEM's generalization and transferability across multiple datasets, including liver, heart, and bone marrow tissues with a pixel size of 4.4 nm, as well as HeLa cell samples with a pixel size of 2.1 nm. }
    \label{fig:3}
\end{figure*}

\subsection{Transfer Learning}
{\bfseries Generalization Capability.}
To evaluate the generalization capability of WaveletEM, we directly tested our model (pre-trained on the mouse cortex dataset) on four distinct electron microscopy datasets without any fine-tuning: bone marrow, heart, liver, and HeLa cells. As shown in Figure~\ref{fig:3}, the pre-trained model demonstrates promising initial performance across these diverse domains, accurately capturing key structural details in most cases. However, subtle domain-specific variations, such as tissue texture and intensity distributions, can still lead to minor discrepancies in reconstruction quality. Overall, the results confirm that WaveletEM retains a notable degree of robustness when exposed to previously unseen EM data, underscoring its potential for broad applicability. This cross-domain generalization ability is a crucial property for biomedical image analysis systems, as real-world deployment often involves data from diverse sources and acquisition settings~\cite{Shen2020Domain}.

{\bfseries Transferability.}
Building on these generalization experiments, we further explored WaveletEM's transfer learning capabilities. Specifically, we used a small subset of training images and set the learning rate to $1\times10^{-5}$, while keeping all other hyperparameters consistent with the original training setup. Each fine-tuning process ran for 1{,}000 iterations. As shown in Table~\ref{table:gengeate}, all metrics demonstrated notable improvements after fine-tuning, with PSNR gains ranging from 0.82 to 2.16 dB, SSIM improvements of 0.03 to 0.06, and LPIPS reductions indicating enhanced perceptual quality. Furthermore, as illustrated in Figure~\ref{fig:3}, this targeted adaptation significantly enhances both quantitative metrics and qualitative performance, leading to sharper details and more faithful reconstructions of domain-specific structures. These improvements highlight the effectiveness of our wavelet-based dual-stream architecture in bridging domain gaps with minimal additional data.

\begin{table}[t]
\centering
\caption{Ablation study of key components in WaveletEM. Best results are in bold.}
\label{table:ablation_waveletem}

\setlength{\tabcolsep}{4.5pt}
\renewcommand{\arraystretch}{1.1}

\begin{tabular}{ccc S[table-format=1.4] S[table-format=1.4]}
\toprule
\shortstack{Wavelet\\Transform} &
\shortstack{Diffusion\\Branch} &
\shortstack{Transformer\\Branch} &
{Resolution Ratio $\uparrow$} &
{LPIPS $\downarrow$} \\
\midrule

$\checkmark$ & $\checkmark$ &            & 0.8581 & 0.0159 \\
$\checkmark$ &            & $\checkmark$ & 0.7342 & 0.0463 \\
           & $\checkmark$ &            & 0.9925 & 0.0164 \\
           &            & $\checkmark$ & 0.7162 & 0.0456 \\
$\checkmark$ & $\checkmark^{\text{high}}$ & $\checkmark^{\text{low}}$  & 0.6365 & 0.0192 \\
$\checkmark$ & $\checkmark^{\text{low}}$  & $\checkmark^{\text{high}}$ & $\mathbf{1.8579}$ & $\mathbf{0.0133}$ \\

\bottomrule
\end{tabular}
\end{table}

\begin{table}[t]
\centering
\caption{Ablation study of sampling steps. Best results in bold. All timing reported as per-image inference time (batch size = 10).}
\label{table:2}

\setlength{\tabcolsep}{4pt}
\renewcommand{\arraystretch}{1.05}

\begin{tabular}{c S[table-format=2.3] S[table-format=1.4] S[table-format=1.4] S[table-format=1.2]}
\toprule
\multirow{2}{*}{Steps} & \multicolumn{4}{c}{Mouse Brain Cortex (Denoise)} \\
\cmidrule(lr){2-5}
& {PSNR $\uparrow$} & {SSIM $\uparrow$} & {LPIPS $\downarrow$} & {Time (s/img)} \\
\midrule
800  & 27.583 & 0.5951 & 0.0155 & 1.79 \\
900  & 27.616 & 0.5949 & 0.0156 & 1.98 \\
1000 & {\bfseries 28.327} & {\bfseries 0.6264} & {\bfseries 0.0133} & 2.18 \\
1500 & 27.625 & 0.5949 & 0.0154 & 3.20 \\
2000 & 27.615 & 0.5952 & 0.0151 & 4.26 \\
\bottomrule
\end{tabular}
\end{table}

\begin{table}[t]
\centering
\caption{Ablation study of loss terms. Best results in bold.}
\label{table:3}

\setlength{\tabcolsep}{3.5pt}
\renewcommand{\arraystretch}{1.05}

\begin{tabular}{l cc cc S[table-format=2.3] S[table-format=1.4] S[table-format=1.4]}
\toprule
& \multicolumn{2}{c}{LFCDM} & \multicolumn{2}{c}{HFTBM} & \multicolumn{3}{c}{Mouse Brain Cortex} \\
\cmidrule(lr){2-3} \cmidrule(lr){4-5} \cmidrule(lr){6-8}
Method & L1 & L2 & L1 & L2 & {PSNR} & {SSIM} & {LPIPS} \\
\midrule

WaveletEM &  & $\checkmark$ &  & $\checkmark$ & 26.951 & 0.5944 & 0.0182 \\
WaveletEM & $\checkmark$ &  &  & $\checkmark$ & 27.554 & 0.5967 & 0.0162 \\
WaveletEM & $\checkmark$ &  & $\checkmark$ &  & 26.649 & 0.5760 & 0.0161 \\
WaveletEM &  & $\checkmark$ & $\checkmark$ &  & {\bfseries 28.327} & {\bfseries 0.6264} & {\bfseries 0.0133} \\

\bottomrule
\end{tabular}
\end{table}

\begin{figure}[h]
    \centering
    \includegraphics[width=0.8\linewidth]{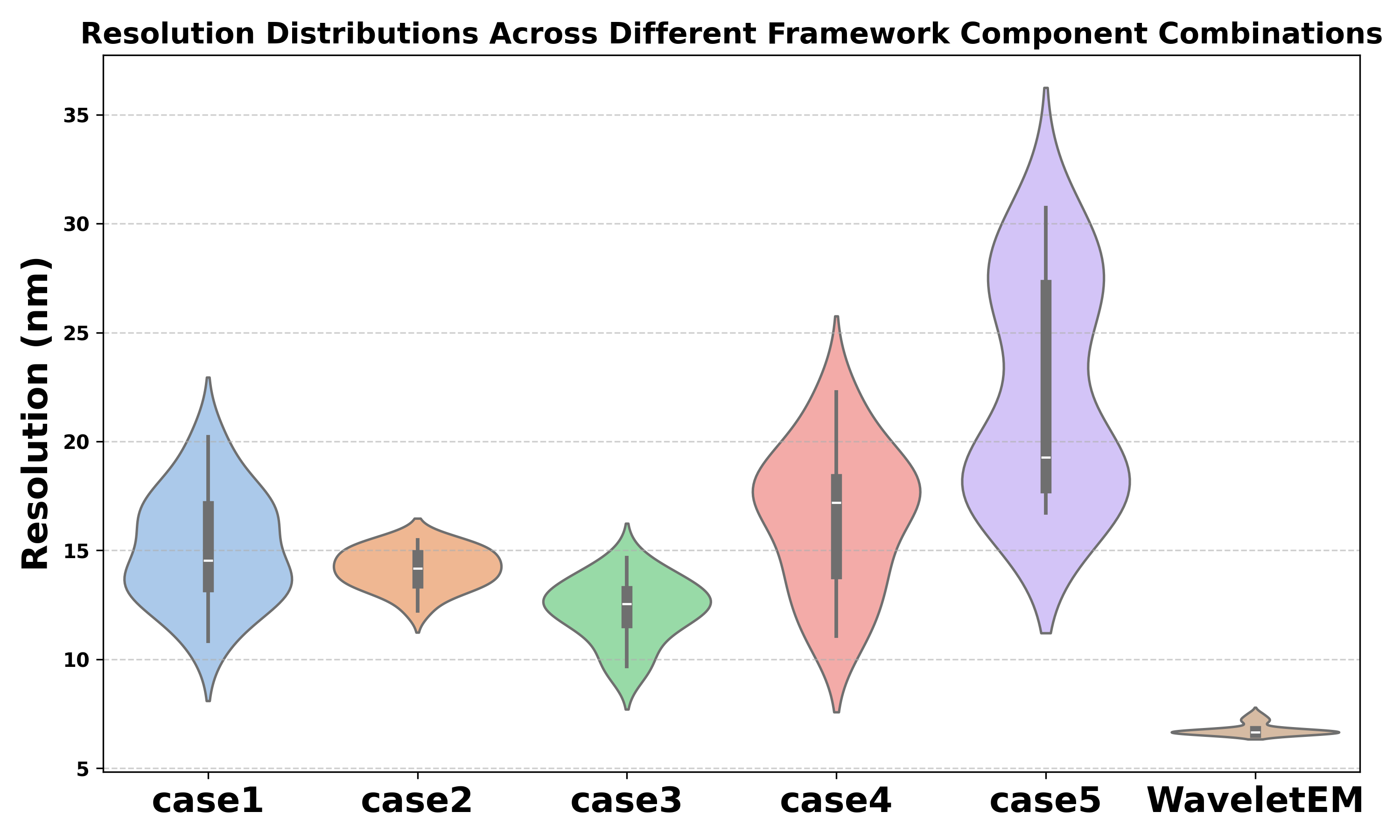}
    \caption{Visualization of Resolution Distributions Across Architectural Component Variants. Case1-5 represents the first five combinations in Table~\ref{table:ablation_waveletem}.}
    \label{fig:ablation_component}
\end{figure}

\begin{figure}[]
    \centering
    \includegraphics[width=0.9\linewidth]{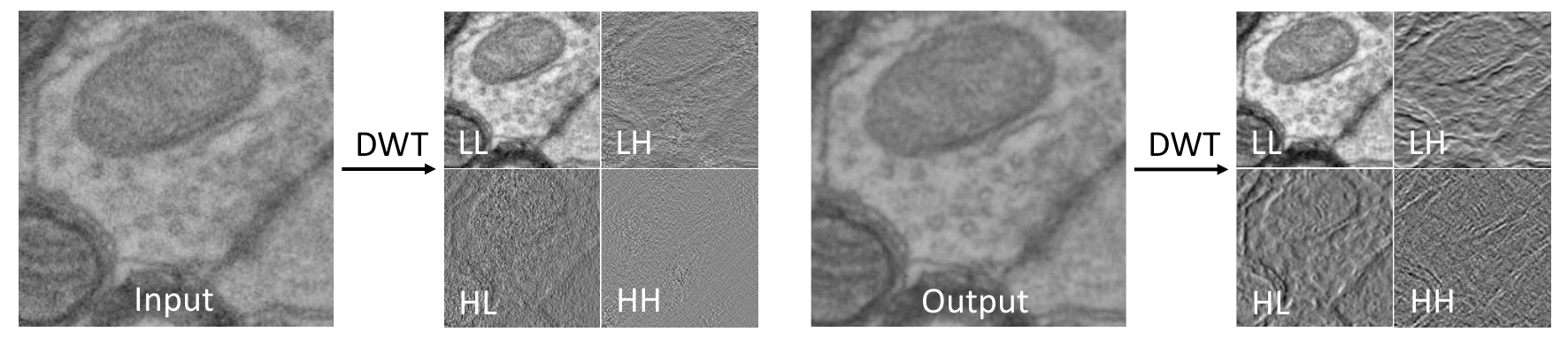}
    \caption{Visualization of the input and output wavelet transforms at sampling step 1000.}
    \label{fig:4}
\end{figure}
\begin{figure}[]
    \centering
    \includegraphics[width=0.9\linewidth]{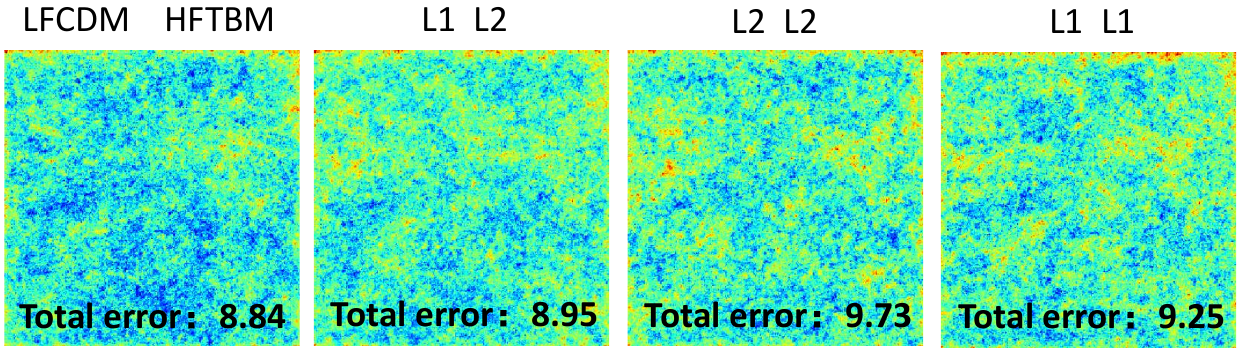}
    \caption{Visualization of error maps for different loss function combinations on the test dataset. The bottom row represents the total error.}
    \label{fig:5}
\end{figure}

\subsection{Ablation Study}
{\bfseries Wavelet Basis. }
We first validate the choice of Haar wavelet by comparing it against two smoother alternatives: Daubechies (db4) and Biorthogonal (bior2.2) wavelets. As shown in Table~\ref{table:wav_ablation}, Haar consistently outperforms both alternatives across all metrics, achieving the highest PSNR (28.327 dB) and SSIM (0.6264) as well as the lowest LPIPS (0.0133). This empirical result aligns with our analysis in Section~\ref{sec2}: the orthogonality and non-redundancy of Haar provide a stable, efficient basis for the independent processing required by our dual-stream architecture, while smoother bases introduce computational overhead without performance gains in this task.

\begin{table}[t]
\centering
\caption{Ablation study on wavelet basis selection. Best results are in \textbf{bold}.}
\label{table:wav_ablation}
\setlength{\tabcolsep}{8pt}
\begin{tabular}{l S[table-format=2.3] S[table-format=1.4] S[table-format=1.4]}
\toprule
Wavelet Basis & {PSNR $\uparrow$} & {SSIM $\uparrow$} & {LPIPS $\downarrow$} \\
\midrule
Daubechies    & 27.456 & 0.5972 & 0.0266 \\
Biorthogonal  & 26.745 & 0.5632 & 0.0247 \\
Haar          & \textbf{28.327} & \textbf{0.6264} & \textbf{0.0133} \\
\bottomrule
\end{tabular}
\end{table}

{\bfseries Architecture Component Analysis. }
To validate the effectiveness of our design, we systematically evaluate six architectural configurations, ranging from single-branch baselines to our complete dual-stream architecture, as detailed in Table~\ref{table:ablation_waveletem}. The results demonstrate that our complete design achieves superior performance, surpassing alternatives by 2-3× in Resolution Ratio while achieving the best (lowest) LPIPS scores. This quantitative superiority is also reflected in the physical resolution limits. Specifically, as illustrated by the resolution distribution in Figure~\ref{fig:ablation_component}, the values achieved by WaveletEM are predominantly concentrated within the 6-8 nm range. In stark contrast, competing methods yield resolutions largely clustered between 10-20 nm, a significantly coarser scale.

{\bfseries Sampling Steps.}
For the Diffusion branch training, we employed 2000 steps. To optimize inference speed while maintaining generation quality, we conducted ablation studies on sampling steps, as shown in Table~\ref{table:2}. Results indicate that 800 and 900 steps yielded insufficient generation quality, while 1500 and 2000 steps showed no significant improvement over 1000 steps despite substantially increased computational time. Furthermore, as illustrated in Figure~\ref{fig:4}, the 1000-step configuration effectively reconstructed all four wavelet subbands, preserving critical frequency information across approximation and detail coefficients. Based on this comprehensive analysis, we selected 1000 sampling steps as the optimal balance between computational efficiency and reconstruction quality for our WaveletEM architecture.

{\bfseries Loss Function.}
To validate the effectiveness of our loss function design, we conducted ablation experiments combining L1 and L2 loss functions across the dual-branch architecture. As demonstrated in Table~\ref{table:3}, applying L2 loss to the diffusion branch while using L1 loss for the transformer branch achieved optimal performance. This configuration strikes an effective balance between the diffusion branch's need for precise structural reconstruction (facilitated by L2's sensitivity to larger errors) and the transformer branch's requirement for preserving fine details and edge information (where L1's robustness to outliers proves beneficial). Furthermore, as illustrated in Figure~\ref{fig:5}, the combination of L2 and L1 loss resulted in the minimal total error compared to other configurations. The experimental results confirm that this loss function combination maximizes the complementary strengths of our dual-stream architecture in electron microscopy image enhancement tasks.

\section{Conclusion}
In this work, we present WaveletEM, a wavelet-based dual-stream architecture that effectively balances realism and fidelity via Frequency-Aware Structural Decomposition. By employing a 2D Discrete Wavelet Transform, WaveletEM decomposes images and processes frequency bands separately. A Low Frequency Conditional Diffusion Model enhances global structure realism, while a High Frequency Transformer-Based Model focuses on reconstructing fine details with high fidelity, achieving computational efficiency and high-quality reconstruction.
Extensive experiments demonstrate WaveletEM's strong performance on the EMDiffuse dataset. It significantly improves image clarity, detail, and resolution while enabling an approximate 60$\times$ acceleration in the computational component of the imaging workflow (replacing the \textasciitilde130 s of conventional long-exposure acquisition with \textasciitilde2.18 s of inference per frame), and showing robust generalization.
WaveletEM advances EM imaging by harmonizing efficiency and precision, enabling faster, reliable workflows in structural biology and nanotechnology.

\section{Limitations and Future Work}

Despite the promising results, several limitations of this study should be acknowledged. 

First, the evaluation relies on image-level quality metrics, which cannot fully capture biological faithfulness. Task-driven validation such as downstream organelle segmentation or expert reader studies would provide stronger evidence of scientific trustworthiness; however, the only publicly available EM dataset currently lacks pixel-level organelle annotations, and establishing such benchmarks remains an important direction for the community. 

Second, the conditional diffusion branch may introduce plausible but false ultrastructural patterns (hallucinations), particularly under extremely low SNR or for structures underrepresented in training. Moreover, due to the limited diversity of publicly available EM datasets—which currently cover only a narrow range of SNR conditions and a few specific pixel sizes—the systematic dependence of model performance on input SNR and pixel size has not been quantitatively characterized.

Third, the current model is trained under a fixed degradation model and a single imaging modality (SEM); its generalization to other microscope types, acquisition protocols, pixel sizes, and inter-laboratory variability has not been systematically evaluated.

Fourth, the per-image inference time (approximately 2.5 seconds) may still not satisfy the latency requirements of real-time or high-throughput microscopy pipelines.

Future work will focus on: (i)~uncertainty-aware reconstruction to signal potential hallucinations; (ii)~user-controllable realism--fidelity trade-offs; (iii)~integration with downstream segmentation or quantitative morphology pipelines; and (iv)~deployment in interactive microscopy workflows.

\section{Code and Data Availability}
To facilitate further research and ensure reproducibility, the complete source code and associated dataset proposed in this study have been made publicly available. The source code can be accessed via GitHub (\url{https://github.com/xiaogaogao26/WaveletEM}) and its corresponding DOI (\url{https://zenodo.org/records/20121164}). The dataset utilized in our experiments is available via DOI (\url{https://zenodo.org/records/10205819}). Detailed information, including environment dependencies and requirements, descriptions and implementations of key algorithms, pretrained model, as well as the recommended citation format, are comprehensively documented in the \texttt{README.md} file within the GitHub repository.


\bibliography{sn-bibliography}

\end{document}